\documentstyle[12pt]{article}
\begin{document}
\begin{center}
$$\;$$
{\Large\bf Quantum  oscillator as 1D anyon}
\end{center}
\begin{center}
Ye. Hakobyan${}^{a)}$,
V. Ter-Antonyan${}^{b)}$
\\[3mm]

      {\it Bogoliubov Laboratory of Theoretical Physics,\\
        Joint Institute for Nuclear Research, \\
        Dubna, Moscow Region, 141980, Russia}
\end{center}

\vspace{2cm}

\begin{center}
\begin{minipage}{12.5cm}

\quad
It is shown that
in one spatial dimension the quantum oscillator is dual to the charged
particle situated in the field described by the  superposition of Coulomb and
Calogero--Sutherland potentials.
\end{minipage}
\end{center}

\vspace{3cm}

{\bf Keywords:}
1D quantum oscillator, reduction,
$1/x^2$ interaction, duality, 1D anyon.

\vfill

\newpage

\noindent
{\large\bf I. INTRODUCTION}

\vspace{0.5cm}

In one spatial dimension a particle moving in the
Calogero--Sutherland potential
$V_{cs}=-\hbar^2\nu(1-\nu)/2\mu x^2$ has a very unusual property.
Unlike the potential $V_{cs}$, the wave function is not invariant
under the replacement $\nu\to (1-\nu)$. It describes a boson for even $\nu$
and a fermion for odd $\nu$. Statistics corresponding to the other values
of $\nu$ is called the fractional statistics${}^{1}$, and the system influenced along with
$V_{cs}$ by a potential binding the particle to the center is called
the 1D anyon${}^{2-4}$. Nobody has observed a 1D anyon yet, but nevertheless
it is of both theoretical${}^{5}$ and experimental${}^6$ interest.
The purpose of the present note is to prove that such an extraordinary object can be
constructed from a 1D quantum oscillator.

\vspace{0.7cm}

\noindent
{\large\bf II. ANYON--OSCILLATOR DUALITY}

\vspace{0.5cm}

Consider the Schr\"odinger equation
\begin{equation}
\label{f1}
\partial_u^2 {\Psi}+\frac{2 \mu}{\hbar^2}\left(
E-\frac{\mu\omega^2 u^2}{2}
\right)\Psi=0,
\end{equation}
which  describes the 1D quantum oscillator. Introduce the quantum number
$s=0, 1/2$ and write $N=2n+2s$, with $N$ numerating the energy levels
$E=\hbar\omega(N+1/2)$ and $n$ being integer and nonnegative. Without
loss of information we can assume $u$ to belong to the region
$0\leq u<\infty$. We interpret $s$ as a spin of the reduced oscillator.
The corresponding wave function is denoted by $\Psi_n^{(s)}$.

Let us look for the function  $\Psi_n^{(s)}$ in the form
\begin{equation}
\label{f2}
\Psi_n^{(s)}(u)=
C\, u^{2 s} \, \overline{\Psi}_n,
\end{equation}
where $\overline{\Psi}_n$ is subordinate to the condition
$\overline{\Psi}_n(0)\not=0$, and $C$ is a normalization constant. Eq.
(\ref{f1}) is easily seen to take the form
\begin{equation}
\label{f3}
\partial_u^2\overline{\Psi}_n+
\frac{4 s}{u}\partial_u \overline{\Psi}_n+
\frac{2 \mu}{\hbar^2}\left(
E-\frac{\mu\omega^2 u^2}{2}
\right)\overline{\Psi}_n=0.
\end{equation}
After change of the variable $x=u^2$,
 we arrive at the equation $(2\nu=2s+1/2)$
\begin{equation}
\label{f4}
\partial_x^2\overline{\Psi}_n+
\frac{2 \nu}{x} \,\partial_x \overline{\Psi}_n+
\frac{2 \mu}{\hbar^2}\left( -\frac{\mu\omega^2}{8}+\frac{E}{4 x}
\right)\overline{\Psi}_n=0.
\end{equation}
Now we set
\begin{equation}
\label{f4a}
\overline{\Psi}_n=x^{-\nu}\Phi_n^{(\nu)},
\end{equation}
then cancel the undesirable term with first derivative in (\ref{f4})
and obtain
\begin{equation}
\label{f5}
\partial_x^2{\Phi_n^{(\nu)}}+
\frac{2 \mu}{\hbar^2}\left( \varepsilon-V_c-V_{cs}
\right){\Phi_n^{(\nu)}}=0,
\end{equation}
where $V_c=-\alpha/x$, $V_{cs}$ is the Calogero--Sutherland
potential with $\nu=1/4$ or $3/4$ and
\begin{equation}
\label{f6}
\varepsilon=-\frac{\mu \omega^2}{8}, \quad \alpha=\frac{E}{4}.
\end{equation}
Eq. (\ref{f5}) describes a system which we call the 1D Coulomb anyon.

This equation realizes a special case of a more general equation
that has a relation to $(2+1)$-dimensional anyons${}^7$.

Comparing Eq. (\ref{f1}) with Eqs. (\ref{f5}) and (\ref{f6}),
we summarize that there are two alternative possibilities connected with
Eq. (\ref{f1}) -- explicit and hidden.
In the first case, the parameter $\omega$ is fixed
($\omega=fix.>0$) and plays a role of coupling constant, the parameter $E$ is quantized
and has a meaning of energy, and the system is a 1D quantum oscillator. For the
hidden possibility, the parameter $E$ is fixed ($E=fix.>0$), the coupling constant
is equal to $E/4$, $\omega$ is quantized, the meaning of energy takes the quantity
$\varepsilon=-\mu\omega^2/8$, and the system is the 1D Coulomb anyon.
In the above-mentioned sense, the 1D quantum oscillator is dual to the 1D Coulomb anyon.

\vspace{0.7cm}

\noindent
{\large\bf III. ENERGY LEVELS AND WAVE FUNCTIONS}

\vspace{0.5cm}

Let us return to Eq. (\ref{f5}) and make the substitution
\begin{equation}
\label{f7}
\Phi_n^{(\nu)}=y^\nu e^{-y/2}
Q(y),
\end{equation}
where $y=x(-8\mu\varepsilon/\hbar^2)^{1/2}$ and $Q(0)\not=0$ and is finite.
The function  $Q(y)$ can diverge at infinity but not higher than the
finite power of $y$. Using (\ref{f7}) and (\ref{f5}) we come to the equation
\begin{equation}
\label{f8}
y\,  \partial_y^2\, Q+(2\nu-y) \partial_y Q-(\nu-\lambda)Q=0,
\end{equation}
with $\lambda=(-\mu\alpha^2/2\hbar^2\varepsilon)^{1/2}$.
Eq. (\ref{f8}) is the equation for a confluent hypergeometric function.
It has a general solution${}^8$
\begin{equation}
\label{f9}
Q(y) = C_1\, F(\nu-\lambda, 2\nu, y)+C_2\, y^{1-2\nu}\, F(1-\lambda-\nu, 2-2\nu, y),
\end{equation}
where $F(a, b, y)$ is given by the series
$$
F(a, b, y)=
1+ \frac{a}{b}\frac{y}{1!}
+\frac{a(a+1)}{b(b+1)}\frac{y^2}{2!} +\dots
$$
convergent for all finite $y$. For large $y$ the asymptotic formula${}^8$
is valid
\begin{equation}
\label{f10}
F(a, b, y)\sim \frac{\Gamma(b)}{\Gamma(b-a)}\, (-y)^{-a}
+ \frac{\Gamma(b)}{\Gamma(a)}\, e^y (y)^{a-b}.
\end{equation}
The second term in (\ref{f9})  for $\nu=3/4$ is singular at $y=0$,
and hence $C_2$ has to be taken zero. The first term in (\ref{f9}), as it is evident
from (\ref{f10}), is ``well-behaved" at infinity under the condition $3/4-\lambda=-n$,
where $n$ is an integer number greater or equal to zero. For $\nu=1/4$ both the
terms in (\ref{f9}) are regular at $y=0$, but the satisfactory behavior at infinity
needs the simultaneous requirements $1/4-\lambda=-n$, $3/4-\lambda=-m$, or $n-m=1/2$,
which is impossible. Hence, either $C_1=0$ or $C_2=0$. But for $C_1=0$ the function
$Q(y)$ will become zero at $y=0$. This contradicts the condition
$Q(0)\not=0$, and, therefore, we put $C_2=0$ and $1/4-\lambda=-n$.
Thus, we conclude that $\nu-\lambda=-n$, i.e.,
\begin{equation}
\label{f11}
\varepsilon_n^{(\nu)}=-\frac{\mu\alpha^2}{2\hbar^2(n+\nu)^2},\quad n=0,1,2,\dots
\end{equation}
Returning to the corresponding eigenfunctions, we put
\begin{equation}
\label{f12}
\Phi_n^{(\nu)}=
C_n^{(\nu)} y^\nu e^{-y/2} F(-n, 2\nu, y).
\end{equation}
It is known${}^9$ that
$$
F(-n, 2\nu, y)=\frac{n! \Gamma(2\nu)}{[\Gamma(n+2\nu)]^2}\, L_n^{2\nu-1}(y),
$$
and
$$
\int\limits_{0}^{\infty} e^{-y} y^{2\nu}
[L_n^{2\nu-1}(y)]^2 \, dy=
2\, (n+\nu)\,
\frac{[\Gamma(n+2\nu)]^3}{n!},
$$
where $L_n^{2\nu-1}(y)$ is an associated Laguerre polynomial.
Using this results and taking into account the relation
$$
\left(-\frac{8\mu\varepsilon}{\hbar^2}\right)^{1/4}=\frac1\hbar
\left(\frac{2\mu\alpha}{n+\nu}\right)^{1/2},
$$
we find
$$
C^{(\nu)}=
\frac{\sqrt{\mu\alpha}}{\hbar}
\frac{1}{n+\nu}\,
\frac{1}{\Gamma(2\nu)}\sqrt{\frac{\Gamma(n+2 \nu)}{n!}}.
$$
Summarizing, we write
\begin{equation}
\label{f13}
\Phi_n^{(\nu)}=
\frac{\sqrt{\mu\alpha}}{\hbar}
\frac{1}{n+\nu}\,
\frac{1}{\Gamma(2\nu)}\sqrt{\frac{\Gamma(n+2 \nu)}{n!}}\,
y^\nu e^{-y/2} F(-n, 2\nu, y).
\end{equation}
So, we have two types of the 1D Coulomb anyons with $\nu=1/4$
and $\nu=3/4$. They are dual to reduced oscillators with $s=0$ and
$s=1/2$, respectively.

\vspace{0.7cm}

\noindent
{\large\bf IV. DUALITY FOR SOLUTIONS}

\vspace{0.5cm}

Now we will calculate the energy levels $\varepsilon_n$ and wave functions
$\Phi_n^{(\nu)}$ in another, more straightforward, way. For energy levels we have
$$
\varepsilon=-\frac{\mu\omega^2}{8}=-\frac{\mu}{8}
\left[\frac{E}{2\hbar( n + \nu)}\right]^2
=-\frac{\mu}{8}
\left[\frac{2\alpha}{\hbar(n + \nu)}\right]^2=
-\frac{\mu\alpha^2}{2\hbar^2(n+\nu)^2}.
$$
It follows from Eqs. (\ref{f2}) and (\ref{f4a}) that
$$
\Phi_n^{(\nu)}=\, \frac{1}{C} \, x^{1/4} \Psi_n^{(\nu)}
$$
and, therefore,
$$
\int\limits_{0}^{\infty}
|\Phi_n^{(\nu)}|^2 \,dx=\frac{1}{|C|^2}\int_0^{\infty} x^{1/2}\,
|\Psi_n^{(s)}|^2 \, dx.
$$
The integral in the left-hand side is equal to 1, from which it follows that
$$
|C|^2=2 \int\limits_{-\infty}^{\infty}\, u^2\, |\Psi_N(u)|^2 \,du =
2 \, \overline{u^2}=
\frac{4(n+\nu)\hbar}{\mu\omega},
$$
where $\Psi_N$ is the normalized wave function of a 1D quantum oscillator.
Thus,
\begin{equation}
\label{f14}
\Phi_n^{(\nu)}=
\frac{(-1)^n}{2} \sqrt{\frac{\mu\omega}{\hbar (n+\nu)}}\,
x^{1/4} \Psi_n^{(s)}.
\end{equation}
Remind that according to the theory of quantum oscillator${}^9$,
\begin{equation}
\label{f15}
\Psi_n^{(s)}=\sqrt{2}
\left(\frac{\mu\omega}{\pi\hbar}\right)^{1/4} \frac{1}{2^N N!}
e^{-\mu\omega u^2/2} H_N \left(u \sqrt{\frac{\mu \omega}{\hbar}}\right).
\end{equation}
Further, it is known${}^{10}$ that Hermite polynomials could be expressed in
terms of confluent hypergeometric functions. For our case
\begin{equation}
\label{f16}
H_{2 n+2 s}(\sqrt{y})= (-1)^n \frac{(2n+2s)!}{n!}
(2\sqrt{y})^{2s} F(-n, 2s+1/2, y).
\end{equation}
Using the identification $y=x \mu\omega/\hbar$ and the relations $2s+1/2=2\nu$
and $\mu\omega/\hbar=2\mu\alpha/\hbar^2(n+\nu)$, and taking into account
Eqs. (\ref{f14})-(\ref{f16}) we get
\begin{equation}
\label{f17a}
\Phi_n^{(\nu)}=
\tilde{C}_n^{(\nu)}\, y^\nu e^{-y/2} F(-n, 2\nu, y),
\end{equation}
where
\begin{equation}
\label{f17b}
\tilde{C}_n^{(\nu)}=
\sqrt{\frac{\mu\alpha}{\hbar^2}}\frac{1}{2^{n-\nu+1/4}}
\frac{\sqrt{\Gamma(2n+2\nu+1/2)}}{\pi^{1/4}n! (n+\nu)},
\end{equation}
or more explicitly
$$
\tilde{C}_n^{(1/4)}=\frac{\sqrt{\mu\alpha}}{\hbar}\frac{1}{2^n}
\frac{\sqrt{\Gamma(2n+1)}}{\pi^{1/4}n! (n+1/4)},
$$
$$
\tilde{C}_n^{(3/4)}=\frac{\sqrt{\mu\alpha}}{\hbar}\frac{1}{2^{n-1/2}}
\frac{\sqrt{\Gamma(2n+2)}}{\pi^{1/4}n! (n+3/4)}.
$$
From the duplication formula for a gamma-function
$$
\Gamma(2z)= 2^{2z-1} \pi^{-1/2} \Gamma(z)\Gamma(z+1/2)
$$
and taking into account that  $\Gamma(1/2)=\pi^{1/2}$,
$\Gamma(3/2)=\frac12\pi^{1/2}$, we conclude that
$\tilde{C}_n^{(\nu)}=C_n^{(\nu)}$ and, consequently, Eqs. (\ref{f17a})
and (\ref{f13}) are identical.

\vspace{0.7cm}

\noindent
{\large\bf V. CONCLUSIONS}

\vspace{0.5cm}

{\bf a)} The 1D oscillator has only a discrete energy spectrum and, therefore,
 is a model provided by the property which is known in QCD as confinement.
A particle situated in the confinement potential cannot be removed from
the center and transferred to infinity. On the
other hand, the 1D Coulomb anyon is a system possessing both the discrete and
continuous part in the energy spectrum.  At the same time, it includes $1/x^2$
interaction and, therefore, pretends to be a magnetic monopole in one spatial
 dimension.  All these ideas confirm that our result can be interpreted in the
spirit of the Seiberg--Witten duality${}^{11}$:  The theories with strong
coupling (i.e., including confinement) are equivalent to the theories with
weak coupling (i.e., without confinement) accompanied by magnetic monopoles.
We conclude that the Seiberg--Witten duality has its prototype in 1D quantum mechanics.

{\bf b)} The anyon--oscillator duality is a simple example of a
 more complicated
dyon--oscillator duality${}^{12-23}$. The latter connects
the isotropic oscillator with charge--dyon bound system
(dyon is a hypothetical object which has both
the electric and magnetic charge${}^{24}$). The passage from an oscillator to
a charge--dyon system is realized by non-bijective bilinear
transformations${}^{25}$ (for the mapping of the 1D Coulomb system into the
oscillator refer to${}^{26}$).

{\bf c)}  The wave function (\ref{f12}) of 1D Coulomb anyon can formally be
extended to the region $-\infty<y<0$. Such a continuation is an arbitrary-rule
operation and we choose the following one. First, still being in the region $0<y<\infty$,
we change $y$ in the exponent and confluent hypergeometric function by $|y|$
and remain unchanged the factor $y^\nu$. Then, we extend the expression to the
region $-\infty<y<0$. These steps allow us to get rid of divergence in
the exponent for large negative values of $y$ and conserve the normalization condition
in $-\infty<y<\infty$ by multiplying the function $\Phi_n^{(\nu)}$ by the factor
$1/\sqrt{2}$. The obtained wave function  $\overline{\Phi}_n^{(\nu)}(y)$ satisfies
Eq. (\ref{f5}) in the region $-\infty<y<\infty$
and has the parity $(-1)^\nu$, i.e. describes the 1D anyon${}^4$.

{\bf d)} Eq. (\ref{f5}) for $-\infty<x<\infty$ and $\nu=0$ corresponds to
the so-called 1D hydrogen atom${}^{27}$ (for later references see${}^{28}$)
which has some mysterious properties. For example,
the ground state corresponds to an infinite negative value of the energy and the exited
levels are double degenerated. The reason is that the potential $(-1/|x|)$ is singular
in 1D space and the system is provided by hidden symmetry${}^{29-31}$
and supersymmetry${}^{32,33}$. As it is follows from (\ref{f5}) and
(\ref{f11}), the Calogero--Sutherland potential transforms the 1D hydrogen
atom into two modified atoms with the statistical parameter $\nu=1/4$ and
$\nu=3/4$. This transformation leads to the formation of the ground states
with a finite energy level and remove the problem of degeneracy (replacement
$n\to n+\nu$).

{\bf e)} It is easily to be convinced
that Eq. (\ref{f4}) is identical to the Schr\"odinger
equation with the Hamiltonian
\begin{equation}
\label{ham}
\hat H=\frac{1}{2\mu}\left(-i\hbar \, \partial_x-\frac{e}{c}A
\right)^2-\frac{\alpha}{x}-\frac{\hbar^2}{2\mu}\frac{\nu(1-\nu)}{x^2}
\end{equation}
where $\alpha=e^2$, $A=g/x$, $g=i \nu\hbar c/e$. So, we deal with
a charged particle
moving in the field created by the 1D Coulomb dyon of the electric charge $e$
and purely imaginary magnetic charge $g$.
The Calogero--Sutherland potential gains the
meaning of the Goldhaber term typical of theory
of magnetic monopoles${}^{34,35}$.

Note that the Hamiltonian in (\ref{ham})  is not Hermitian, but it
could be transformed into the Hermitian one if we do the following:
1) consider instead of the semiaxis $x \in (0, \infty)$
the axis $x \in (-\infty, \infty)$;
2) substitute $\alpha/x$ by $\alpha/|x|$;
3) introduce the Yang--Dunkl operator${}^{36}$ $\hat{D}=-i\hbar\partial_x-e
A\hat{R}/c$ for the Calogero model, where $\hat{R}$ is the reflection operator.

Exactly the same ``covariant derivative" appears in the Calogero-like models with exchange
interaction${}^{37,38}$.

\vspace{0.7cm}

\noindent
{\large\bf ACKNOWLEDGMENT}

\vspace{0.5cm}

We thank our collaborators L. Mardoyan, A. Nersessian, G. Pogosyan and A. Sissakian.
Their evident interest in the subject leads us
to the writing of present short note.

Also we are grateful to M.S. Plyushchay for valuable remarks on $(2+1)$-dimensional
anyons and the Yang--Dunkl operator for Calogero models.

The work of Ye. Hakobyan was supported in part by the
Russian Foundation for Basic Research, project no. 98-01-00330.

\vspace{3cm}

\noindent
${}^{a)}$ Electronic mail: yera@thsun1.jinr.ru

\noindent
${}^{b)}$ Electronic mail: terant@thsun1.jinr.ru

\noindent
${}^{1}$ A. P. Balachandran,
``Classical topology and quantum statistics," Int. J. Mod. Phys. B {\bf 5},
2585-2623 (1991).

\noindent
${}^{2}$ A. P. Polychronakos,
``Non-relativistic bosonization and fractional statistics,"
Nucl. Phys. B {\bf 324}, 597-622 (1989).

\noindent
${}^{3}$ S. Isakov, ``Fractional statistics in one dimension: modeling
by means of $1/x^2$ interaction and statistical mechanics,"
Int. J. Mod. Phys. A {\bf 9}, 2563-2582 (1994).

\noindent
${}^{4}$ J. Camboa, J. Zanelli, ``Anyons in 1+1 dimensions," Phys. Lett. B
{\bf 357}, 131-137 (1995).

\noindent
${}^{5}$ Z. H. C.  Ha, ``Fractional statistics in one dimensions:
view from an exactly solvable model," Nucl. Phys. B {\bf 435}, 604-636 (1995).

\noindent
${}^{6}$ C. I. Kane and M. P. A. Fisher,
``Transmission through barrier and resonant tunneling in an interacting
one-dimensional electron gas," Phys. Rev. B {\bf 46}, 15233-15262 (1992).

\noindent
${}^{7}$ M. S. Plyushchay, ``Quantization of the classical $SL(2,R)$
system and representations of $SL(2,R)$ group," J. Math. Phys. {\bf 34},
3954-3963 (1993).

\noindent
${}^{8}$ L. D. Landau and E. M. Lifshitz,
{\it Quantum mechanics: nonrelativistic theory}, (Nauka, Moscow, 1989),
pp. 755-758 (in Russian).

\noindent
${}^{9}$ E. Merzbacher,
{\it Quantum Mechanics} (John Wiley\&Sons, New York-Chichester-Brisbane-Toronto-Singapore,
1970), p. 61, p. 209.

\noindent
${}^{10}$ A. Erdelyi, W. Magnus, F. Oberheittinger, and
F.G. Tricomi,
{\it Higher transcendental functions} (McGraw --Hill Book Co., New York, 1953)
v. 2, p. 194.

\noindent
${}^{11}$ N. Seiberg, E. Witten,
``Monopoles, duality and chiral symmetry breaking
in $N=2$ supersymmetric QCD," Nucl. Phys. B {\bf 431}, 484-550 (1994).

\noindent
${}^{12}$ A. P. Chen and M. Kibler, ``Connection between the hydrogen atom
and the four-dimensional oscillator," Phys. Rev. A {\bf 31}, 3960-3963 (1985).

\noindent
${}^{13}$  Toshiniro Iwai, ``The geometry of the SU(2) Kepler problem,"
JGP-{\bf 7}, 507-535 (1990).

\noindent
${}^{14}$ M. Trunk, ``The five-dimensional Kepler problem as an SU(2)
gauge system: algebraic constraint quantization,"
Int. J. Mod. Phys. A {\bf 11}, 2329-2355 (1996).

\noindent
${}^{15}$ I. Mladenov and V. Tsanov, ``Geometric quantization of the
MIC-Kepler problem," J. Phys. A {\bf 20}, 5865-5871 (1987).

\noindent
${}^{16}$  M. V. Pletyukhov and E. A. Tolkachev,
``8D oscillator and 5D Kepler problem: the case of nontrivial constraints,"
J. Math. Phys. {\bf 40}, 93-100 (1999).

\noindent
${}^{17}$  L. Davtyan, L. Mardoyan, G. Pogosyan, A. Sissakian,
V. Ter-Antonyan,
``Generalized KS transformation: from five-dimensional hydrogen atom
to eight-dimensional isotropic oscillator,"
J. Phys. A {\bf 20}, 6121-6125 (1987).

\noindent
${}^{18}$  A. Maghakian, A. Sissakian, V. Ter-Antonyan,
``Electromagnetic duality for anyons," Phys. Lett. A {\bf 236}, 5-7 (1997).

\noindent
${}^{19}$  A. Nersessian, V. Ter-Antonyan, M. Tsulaia,
``A note on quantum Bohlin transformation," Mod. Phys. Lett. A {\bf 11},
1605-1610 (1996).

\noindent
${}^{20}$  V. Ter-Antonyan, A. Nersessian, ``Quantum oscillator and
a bound system of two dyons," Mod. Phys. Lett. A {\bf 10}, 2633-2638 (1995).

\noindent
${}^{21}$  L. Mardoyan, A. Sissakian, V. Ter-Antonyan, ``8D
oscillator as a hidden SU(2) monopole," Phys. Atom. Nucl.
{\bf 61}, 1746-1750 (1998).

\noindent
${}^{22}$ L. Mardoyan, A. Sissakian, V. Ter-Antonyan,
``Hidden symmetry of Yang-Coulomb monopole,"
Mod. Phys. Lett. A {\bf 14}, 1303- 1307 (1999).

\noindent
${}^{23}$ A. Nersessian, V. Ter-Antonyan,
``Anyons, monopoles and Coulomb problem,"
Phys. Atom. Nucl. {\bf 61}, 1756-1761 (1998).

\noindent
${}^{24}$ J. Schwinger,
``Magnetic model of matter,"
Science {\bf 165}, 757-761 (1969).

\noindent
${}^{25}$  D. Lambert, M. Kibler, ``An algebraic and geometric approach to
non-bijective quadratic transformations," J. Phys. A {\bf 21}, 307-343 (1988).

\noindent
${}^{26}$  D. S. Bateman, C. Boyd, B. Dutta-Roy,
``The mapping of the Coulomb problem into the oscillator,"
Amer. J. Phys. {\bf 60}, 833-836 (1992).

\noindent
${}^{27}$  R. Loudon, ``One-dimensional hydrogen atom,"
Amer. J. Phys. {\bf 27}, 649-655 (1959).

\noindent
${}^{28}$  U. Oseguera, M. de Llano, ``Two singular potentials: the
space-splitting effect," J. Mat. Phys. {\bf 34}, 4575-4589 (1993).

\noindent
${}^{29}$ L. Davtyan, G. Pogosyan, A. Sissakian, V. Ter-Antonyan,
``On the hidden symmetry of a one-dimensional hydrogen atom," J. Phys. A
{\bf 20}, 2765-2772 (1987).

\noindent
${}^{30}$ L. Boya, M. Kmiecik, A. Bohm,
``Hydrogen atom in one dimension," Phys. Rev. A {\bf 37}, 3567-3569 (1988).

\noindent
${}^{31}$ T. A. Weber, C. L. Hammer,
``The one-dimensional Coulomb potential as a generalized function and the hidden
O(2) symmetry," J. Math. Phys. {\bf 31}, 1441-1444 (1990).

\noindent
${}^{32}$ A. Sissakian, V. Ter-Antonyan, G. Pogosyan, I. Lutsenko,
``Supersymmetry of one-dimensional hydrogen atom,"
Phys. Lett. A {\bf 143}, 247-249 (1990).

\noindent
${}^{33}$ H. N. N\'u\~nez Y\'epez, C. A. Vargas,
``Superselection rule in the one-dimensional hydrogen atom,"
J. Phys. A {\bf 21}, L651-L653 (1988).

\noindent
${}^{34}$  A. Goldhaber,
``Role of spin in monopole problem,"
Phys. Rev. B  {\bf 140}, 1407-1414 (1965).

\noindent
${}^{35}$  D. Zwanziger,
``Exactly soluble nonrelativistic model of particles with both electric and magnetic
charge," Phys. Rev. {\bf 176}, 1480-1488 (1968).

\noindent
${}^{36}$   L. M. Yang, ``A note on the quantum rule of the harmonic
oscillator," Phys. Rev. {\bf 84}, 788-790 (1951).

\noindent
${}^{37}$   J. Gamboa, M. Plyushchay, J. Zanelli, ``Three aspects of
bosonized supersymmetry and linear differential field equation with
reflection," Nucl. Phys. {\bf B 543}, 447-465 (1999).

\noindent
${}^{38}$   M. Plyushchay, ``Hidden nonlinear supersymmetries on pure
parabosonic system," hep-th/9903130.

\end{document}